%% file: Lattice2017_203_PARK.tex
\documentclass[epj]{webofc}
\usepackage[utf8]{inputenc}
\usepackage[varg]{txfonts}   % Web of Conferences font
\usepackage{booktabs}
\usepackage{xcolor}
\definecolor{darkred}{rgb}{0.4,0.0,0.0}
\definecolor{darkgreen}{rgb}{0.0,0.4,0.0}
\definecolor{darkblue}{rgb}{0.0,0.0,0.4}
\usepackage[bookmarks,linktocpage,colorlinks,
    linkcolor = darkred,
    urlcolor  = darkblue,
    citecolor = darkgreen]{hyperref}
\usepackage{subfigure}
\usepackage{multirow}
\usepackage{caption}
\wocname{EPJ Web of Conferences}
\woctitle{Lattice2017}
\graphicspath{{./figs}}

\input{macro}

\allowdisplaybreaks
\begin{document}
%%%%%%%%%%%%%%%%%%%%%%%%%%%%%%%%%%%%%%%%%%%%%%%%%%%%%%%%%%%%%%%%%%%%%%%%%%%%%
%
\selectlanguage{english}
%----------------------------------------------------------------------------
\title{ Calculation of $\bar B \rightarrow D^\ast \ell \bar \nu$ form
  factor at zero recoil using the Oktay-Kronfeld action }
%----------------------------------------------------------------------------
\author{%
\firstname{Jon A.} \lastname{Bailey}\inst{1} \and
\firstname{Tanmoy} \lastname{Bhattacharya}\inst{2} \and
\firstname{Rajan} \lastname{Gupta}\inst{2} \and
\firstname{Yong-Chull} \lastname{Jang}\inst{2}\fnsep\inst{3} \and
\firstname{Weonjong} \lastname{Lee}\inst{1}\fnsep\thanks{\email{wlee@snu.ac.kr}} 
\and
\firstname{Jaehoon} \lastname{Leem}\inst{1} \and
\firstname{Sungwoo}  \lastname{Park}\inst{1}\fnsep\thanks{Speaker} \and
\firstname{Boram} \lastname{Yoon}\inst{4} \and
\lastname{(LANL-SWME Collaboration)}
}
%----------------------------------------------------------------------------
\institute{
  Lattice Gauge Theory Research Center, CTP,
  Department of Physics and Astronomy,\\
  Seoul National University, Seoul 08826, South Korea
  \and
  Los Alamos National Laboratory,
  Theoretical Division T-2,
  Los Alamos, New Mexico 87545, USA
  \and
  Brookhaven National Laboratory,
  Department of Physics, Upton, New York 11973, USA
  \and
  Los Alamos National Laboratory,
  Computer, Computational, Statistical Science Division CCS-7,\\
  Los Alamos, New Mexico 87545, USA
}
%----------------------------------------------------------------------------
\abstract{ \rg{We present the first preliminary results for the
    semileptonic form factor $h_{A_1}(w=1)/\rho_{A_j}$} at zero recoil
  for the $\bar B \rightarrow D^\ast \ell \bar \nu$ decay using
  lattice QCD with four flavors of sea quarks. We use the HISQ
  staggered action for the light valence and sea quarks (the MILC HISQ
  configurations), and the Oktay-Kronfeld (OK) action for the heavy
  valence quarks.}
%----------------------------------------------------------------------------
\maketitle
%----------------------------------------------------------------------------
\section{Introduction}
\label{sec:intro}

\rg{The $4.1\sigma$ tension between recently updated values of the
exclusive and inclusive $|V_{cb}|$ \cite{Amhis:2016xyh,Lee:2017proc}, 
and the $4.0\sigma$ tension between the standard
model prediction of $\epsK$ using the exclusive $|V_{cb}|$
determined with lattice QCD inputs and the experimental value of
$\epsK$~\cite{Lee:2017proc} 
motivates highter precision lattice calculations of $|V_{cb}|$.}\looseness-1

\rg{At present, the largest error in exclusive $|V_{cb}|$ comes from the
about $1\%$ heavy quark discretization error in 
the Fermilab action used in current calculations~\cite{ ElKhadra:1996mp}. 
To reduce this error, we are simulating the Oktay-Kronfeld
action~\cite{ Oktay:2008ex}, which is an $\mathcal{O}(\lambda^3)$
improved version of the Fermilab action~\footnote{In the HQET power
  counting, $\lambda \approx {\Lambda}/{(2 m_Q)}$ where $\Lambda
  \approx 300 \MeV$ and $m_Q$ is the heavy quark mass. For the charm
  quark, $\lambda \approx 1/8$ and for the bottom quark, $\lambda
  \approx 1/30$.}.
In this talk, we present the first results on
the semileptonic form factor $h_{A_1}(w=1)/\rho_{A_j}$ at zero recoil
for the $\BtoDst$ decay. The calculations were done on the $a12m310$ 
2+1+1-flavor HISQ lattices generated by the MILC collaboration~\cite{
  Bazavov:2012xda}.}

%----------------------------------------------------------------------------
\section{Nonperturbative determination of $\kcrit$, $\kappa_c$ and $\kappa_b$}
\label{sec:kcrit}

The Oktay-Kronfeld (OK) action \cite{ Oktay:2008ex} is a highly
improved version of the Fermilab action.
\rg{It contains counter-terms up to $\lambda^3$, whereas 
the Fermilab action is improved only up to $\lambda^1$ order.
In both actions, the bare quark mass is defined as}
\begin{align}
  am_0 = \frac{1}{2u_0} \left( \frac{1}{\kappa} -
  \frac{1}{\kcrit}\right) \,,
\end{align} 
\rg{where $u_0$ is the tadpole improvement factor determined from the
$1 \times 1$  Wilson loops.
In case of the Fermilab action, the coefficients of the dimension 5
counter terms added for improvement are independent of $m_0$. It is therefore 
straight-forward to determine $\kcrit$ by fitting the pion
spectra with respect to $am_0$.
In case of the OK action, some of the coefficients of the dimension 5 and
dimension 6 counter terms, such as $c_E$, $c_1$, $c_2=c_3$, $c_{EE}$
defined in~\cite{ Oktay:2008ex}, depend on $m_0$ and $\kcrit$. 
Hence, for the OK action we determine $\kcrit$ by iteration as described below.}
\begin{enumerate}
\item First, we determine $\kcrit^\tree$ using the following formula:
  \begin{align}
    \kcrit^\tree &= \dfrac{1}{2u_0\cdot (1+3\zeta r_s + 18c_4)}
    = 0.053850\cdots
  \end{align}
where we set $\zeta = r_s = 1$ and $c_4$ is given in
Ref.~\cite{ Oktay:2008ex}. \rg{For the HISQ ensemble with 
$a \approx 0.12\fm$ and $M_\pi = 310$~MeV, $a12m310$, the quark 
masses are $\{am_\ell, am_s, am_c\} = \{0.0102,0.0509,0.635\}$.}
\item Make an initial guess for $\kcrit$. We choose
  $\kcrit'=0.96\kcrit^\tree$ \rg{based on Ref.~\cite{Bernard:2010fr}, where it was shown 
  that the nonperturbative   $\kcrit^\text{NP}$ is 4\% smaller than 
$\kcrit^\tree$ for the   Fermilab action.}
\item\label{item:1} Determine the OK action coefficients using
  $\kcrit'$ \rg{and} measure 2-point pion correlators with point sources and
  sinks. We use 3 different $\kappa$ values such that pions have
  masses in the range of $600 \le m_\pi \le 950\MeV$. \rg{Also, we investigate 
  9 different pion momenta $\mathbf{p}$.}
\item \rg{We determine the pion kinetic mass, $M_2$, by fitting the
  ground state energy, $E({\bf p})$, to the dispersion relation}
  \begin{align}
    E(\mathbf{ p}) = M_1 + \frac{\mathbf{p}^2}{2 M_2} -
  \frac{(\mathbf{p}^2)^2}{8M_4^3}-\frac{a^3W_4}{6}\sum_{i=1}^3 {p_i^4}.
  \label{eq:disp}
  \end{align}
  Here $M_1$ is the rest mass, $M_4$ is the quartic mass and $W_4$
  is the Lorentz symmetry breaking term.
\item Determine \rg{the new} $\kcrit$ by requiring $A=0$ in the quadratic fit
  to the following function:
  \begin{align}
      M_2^2(\kappa,\kcrit') &= A+B
      m_2(\kappa,\kcrit)+Cm_2^2(\kappa,\kcrit)\label{eq:kcrit_fit_quad}
  \end{align}
  where the kinetic quark mass $m_2$ is related to the $m_0$ by the
  tree-level relation:
  \begin{align}
      \frac{1}{a m_2} &= \frac{2 \zeta^2}{ am_0 (2+ am_0)} + \frac{
        r_s \zeta }{1+ am_0},
  \label{eq:m2_m0}
  \end{align}
  Since $m_2(\kcrit,\kcrit)=0$, $M_2(\kappa,\kcrit')$ vanishes
  at $\kappa = \kcrit$.
\item Update $\kcrit'=\kcrit$ and \rg{go back to} step \ref{item:1}.
\end{enumerate}

\input{fig_kcrit}

\rg{Fig.~\ref{fig:kcrit} shows the convergence of $\kcrit$ as a function
of the iteration number. 
After the two iterations, we declare convergence to $\kcrit^\text{NP} =
0.051211(33)$ within statistical uncertainty. 
Further details  will be reported in Ref.~\cite{Park:2017}.}
%

% kappa_c, kappa_b
%
We measure the two point correlation functions for the $B_s$ and $D_s$
mesons (pseudo-scalar channel) using the OK action with \rg{the above} nonperturbatively
determined $\kcrit = \kcrit^\text{NP}$. \rg{To determine $\kappa_c$ and $\kappa_b$, we 
simulate at four values of $\kappa$ about estimates for the charm and
the bottom quarks.
For each $\kappa$, we obtain data for 11 momenta and determine the kinetic masses by
fitting to the dispersion relation in Eq.~\eqref{eq:disp}. The final values of 
$\kappa_c$ and $\kappa_b$, obtained by requiring the 
meson masses match the experimental $B_s$ and $D_s$ masses, 
are given in Table \ref{tab:kc,kb}, for two determinations of the 
lattice spacing $a$, $a_{f_{\pi^+}}$ and $a_{r_1}$. Matching the pseudoscalar masses 
gives estimates with the smallest uncertainty.}

\input{tab_kc_kb}

%

%%%%%%%%%%%%%%%%%%%%%%%%%%%
% Section
%%%%%%%%%%%%%%%%%%%%%%%%%%%
\section{Inconsistency \label{sec:incon}}

The inconsistency \rg{parameter $I$} is defined as
\begin{align}
  I &\equiv
  \frac{2\delta M_{\wbar{Q}q} - ({\delta}M_{\wbar{Q}Q} + {\delta}M_{\wbar{q}q})}
       {2M_{2\wbar{Q}q}}
= \frac{2{\delta}B_{\wbar{Q}q} - ({\delta}B_{\wbar{Q}Q} + {\delta}B_{\wbar{q}q})}
       {2M_{2\wbar{Q}q}},
  \label{eq:iparam}
\end{align}
where $\delta M_{X} \equiv M_{2X} - M_{1X}, (X=\wbar{Q}q, \wbar{Q}Q)$.
In a relativistically invariant theory, the binding energy $B_1$ is equal
to $B_2$.
The inconsistency $I$ probes the binding energy difference $\delta B =
B_2 - B_1$.
This difference comes from the discretization errors in the
$\mathcal{O}((ap)^4)$ terms of the action or in the $\mathcal{O}(v^4)$
in NRQCD power counting.
\rg{$I$} vanishes at tree level for the OK action,
but not for the Fermilab action \cite{ Kronfeld:1996uy, Bernard:2010fr}.

In Fig.~\ref{fig:I}, \rg{we show $I$} as a function of
the pseudo-scalar heavy-light meson mass.
Our previous results presented in Refs.~\cite{ Bailey:2017nzm,
  Bailey:2014zma, Bailey:2016kbw} were obtained using the tree-level
$\kcrit^\tree$ with asqtad strange quarks with point sources, while
the results presented in Fig.~\ref{fig:I} are obtained using the
nonperturbatively determined $\kcrit^\text{NP}$ with HISQ strange
quarks with covariant Gaussian smearing applied to the heavy quarks.
We find that the inconsistency parameter vanishes within statistical
uncertainty near the $B_s$ region and it is smaller than that of
Fermilab action by order of magnitude near the $D_s$ region.
\rg{This improvement is due to the combination of using the OK action on HISQ
  ensembles and smeared sources. It is observed for both
  $\kcrit^\tree$ and $\kcrit^\text{NP}$.}
%

\input{fig_I}

%

%%%%%%%%%%%%%%%%%%%%%%%%%%%
% Section
%%%%%%%%%%%%%%%%%%%%%%%%%%%
\section{Form factor $h_{A_1}(1)/\rho_{A_j}$ at zero recoil}
\label{sec:hA1}

To extract the form factor $h_{A_1}(1)$ at zero recoil, we calculate
the double ratio $R$ on the lattice \cite{ Hashimoto:2001nb,
  Bernard:2008dn, Bailey:2014tva}:
\begin{align}
  R(t,t_f) &\equiv \frac{C^{B\rightarrow
      D^\ast}_{A_1}(t,t_f)C^{D^\ast\rightarrow B}_{A_1}(t,t_f)}{
    C^{B\rightarrow B}_{V_4}(t,t_f)C^{D^\ast\rightarrow
      D^\ast}_{V_4}(t,t_f)} \quad
   \xrightarrow[t \to \infty]{t_f \to \infty} \quad
   \left| \frac{h_{A_1}(1)}{\rho_{A_j}}\right|^2 \quad
   \xrightarrow[a \to 0]{V \to \infty} \quad |h_{A_1}(1)|^2 \,.
  \label{eq:R:h_A1}
\end{align}
Here, $\rho_{A_j}$ is the matching factor at $a \ne 0$. 
$C_{J_\mu}^{X \to Y}(t,t_f)$ is a 3-point correlation function:
for example, if $X = B$, $Y = D^\ast$ and $J_\mu = A_j$,
\begin{align}
  C^{B\rightarrow D^{\ast}}_{A_j} (t,t_f)&=\sum_{\mathbf x, \mathbf y}
  \langle O^{D^\ast}_j(0)^\dagger A_j^{cb}(\mathbf
  y,t)O^{B}(\mathbf{x},t_f) \rangle
  \label{eq:C_A1}
\end{align}
\rg{We define the axial and vector currents as follows:}
\begin{align}
  A^{cb}_j &= \bar{\Psi}^c \gamma_j \gamma_5 \Psi^b,
  \qquad V_4^{bb}=\bar{\Psi}^b \gamma_4 \Psi^b
  \\
  \Psi (x)&=\sum_{m=0} d_m \CR_m \psi(x)
  \nonumber \\
  &=\Big[ 1 + {d_1}a
    \boldsymbol{\gamma \cdot D} +  {d_2}a^2
    \Delta^{(3)} + {d_B}a^2  i\boldsymbol{\Sigma\cdot
      B}  -  {d_E}a^2 \boldsymbol{\alpha\cdot E}+
     {d_{rE}}a^3\{ \boldsymbol{\gamma \cdot D},
    \boldsymbol{\alpha\cdot E}\}  \nonumber \\ &\quad -{d_3}a^3\sum_i \gamma_i
    D_i\Delta_i - {d_4}a^3 \{ \boldsymbol{\gamma \cdot D}, \Delta^{(3)}\}
    -  {d_5}a^3 \{ \boldsymbol{\gamma \cdot D}, i\boldsymbol{\Sigma\cdot
      B}\} \nonumber \\
&\quad +  {d_{EE}}a^3\{\gamma_4 D_4, \boldsymbol{\alpha\cdot
      E}\} -  {d_6}a^3 [ \gamma_4 D_4, \Delta^{(3)} ]
    -  {d_{7}}a^3 [\gamma_4 D_4, i\boldsymbol{\Sigma\cdot B} ] \Big]\psi(x).
  \label{eq:rot}
\end{align}
where $d_0=1$, $\CR_0=1$, $\CR_1 = a \boldsymbol{\gamma \cdot D}$,
and so on.
Here, $\Psi$ and $\psi$ are improved and unimproved quark fields,
respectively, \rg{and} the coefficients $d_m$ are real-valued analytic function of the
bare mass $am_0$ \cite{ Leem:2017proc}.
The currents are improved up to order $\lambda^3$ at the tree level
in the HQET power counting.
We obtain the improved fields $\Psi$ using the field rotation defined in
Eq.~\eqref{eq:rot}.
\rg{In short, using} the field rotation is sufficient for the current
improvement at the tree level \cite{ Bailey:2014jga, Bailey:2016wza,
  Leem:2017proc}.

We can rewrite the currents as follows,
\begin{align}
  A^{cb}_j &= \bar{\Psi}^c \gamma_j \gamma_5 \Psi^b
  = \sum_{m,n=0} d_m(m_0^c) \; d_n(m_0^b) \; \bar{A}^{cb}_{j,mn}
  \\
  \bar{A}^{cb}_{j,mn} &\equiv \bar{\psi}^c \CR_m^\dagger
  \gamma_j \gamma_5 \CR_n \psi^b
\end{align}
\rg{and $C^{B\rightarrow D^{\ast}}_{A_j} (t,t_f)$ as}
\begin{align}
  C^{B\to D^\ast}_{A_j} (t,t_f) &=
  \sum_{m,n=0}d_m(m_0^c) \; d_n(m_0^b) \; \bar{C}^{B\rightarrow
    D^\ast}_{A_j;mn} (t,t_f)
  \label{eq:C_A1}
  \\
  \bar{C}^{B\to D^\ast}_{A_j;mn} (t,t_f) &\equiv
  \sum_{\mathbf
    x \mathbf y } \langle O_j^{D^\ast}(0)^\dagger
  \bar{A}^{cb}_{j,mn}(\mathbf{y},t)  O^{B}(\mathbf{x},t_f) \rangle.
\end{align}

There are 12 different rotation operators for $\psi^b$ and $\bar\psi^c$.
In total sum, we have $144 = 12^2$ terms.
However, some of them are $\mathcal{O}(\lambda^4)$, which we exclude
in this analysis.
Hence, we end up with 30 terms up to $\mathcal{O}(\lambda^3)$.
Other 3-point functions \rg{occuring} in Eq.~\eqref{eq:R:h_A1} can be rewritten in
a similar way.

In the measurements, we use point sources \rg{for the HISQ light quarks and choose their coordinates randomly.}
For the $c$ and $b$ quarks, we apply the covariant Gaussian smearing
with $(\sigma,N_\text{GS})=(1.5,5)$ at the source and sink
\rg{points}~\cite{ Yoon:2016dij}.
We also use the coherent source method \cite{ Yoon:2016dij} which gives a
statistical gain by factor of 3 in the measurements.
In the next subsections, we \rg{describe the extraction of $R^{1/2}=h_{A_1}/\rho_{A_j}$
using two different analyses.}

\subsection{Direct analysis on 3-point functions
  $C^{B\rightarrow D^\ast}_{A_1}$ }
\label{sec:fitC}

The fitting function \rg{used for analyzing} $C^{B\rightarrow D^\ast}_{A_1}$ is
\begin{align}
  C^{B\rightarrow D^\ast}_{A_1}
  (t,t_f)&=B^{B\rightarrow D^\ast}e^{-(M_D^\ast-M_B)t}e^{-M_Bt_f}
  (1+\hat{c}^{B\rightarrow D^\ast}(t,t_f))\\
  \bar{C}^{B\rightarrow D^\ast}_{A_1;mn}
  (t,t_f)&=B_{mn}^{B\rightarrow D^\ast}e^{-(M_D^\ast-M_B)t}e^{-M_Bt_f}
  (1+\hat{c}^{B\rightarrow D^\ast}_{mn}(t,t_f)),
\end{align}
where $B^{B\rightarrow D^\ast} = \langle D^\ast | A^{cb}_1 |
B\rangle$, $B^{B\rightarrow D^\ast}_{mn} = \langle D^\ast |
\bar{A}^{cb}_{1,mn} | B\rangle$, and $\hat{c}^{B\rightarrow D^\ast}$
and $\hat{c}^{B\rightarrow D^\ast}_{mn}$ represent the contamination
from the excited states of $B$ and $D^\ast$ mesons.
Once we extract \rg{the $B$ parameters from the data}, we can obtain
$R$ using the following relation:
\begin{align}
  R=\frac{B^{B\rightarrow D^\ast} \cdot B^{D^\ast\rightarrow B}}{
    B^{B\rightarrow B} \cdot B^{D^\ast\rightarrow D^\ast}} =
  \left|\frac{h_{A_1}(1)}{\rho_{A_j}}\right|^2
\end{align}
The amplitude $B$ of 3-point correlation function with
$O(\lambda^p)$ improved current is
\begin{align}
  B = B_{00} + \sum_{(m,n) \ne (0,0)} d_m(m_0^c)\, d_n(m_0^b)\; B_{mn}
  = B_{00}\left[1
    + \sum_{(m,n) \ne (0,0)} d_m(m_0^c) \, d_n(m_0^b) \;
    \frac{B_{mn}}{B_{00}} \right]\,.
  \label{eq:B}
\end{align}
\rg{The} merit of Eq.~\eqref{eq:B} is that the ratio
of $B_{mn}/B_{00}$ can be obtained by a simple constant fit
to the following ratio:
\begin{align}
   {F}_{mn}(t,t_f) &\equiv \frac{ \bar{C}^{B\rightarrow
       D^\ast}_{A_j;mn} (t,t_f)}{ \bar{C}^{B\rightarrow
       D^\ast}_{A_j;00} (t,t_f)} =
   \frac{B_{mn}}{B_{00}} \Big[ 1+\hat{c}^{B\rightarrow
     D^\ast}_{mn}(t,t_f)-\hat{c}^{B\rightarrow
       D^\ast}_{00}(t,t_f) + \cdots \Big]\,.
   \label{eq:F}
\end{align}
\rg{We find that the corrections to the leading term, contamination from the excited states, 
are small and under control.}
In fact, we can further reduce the contamination \rg{using} the
following linear combination \cite{ Bernard:2008dn, Bailey:2014tva}:
\begin{align}
  \bar{F}_{mn} (t,t_f) &\equiv \frac{1}{2} {F}_{mn} (t,t_f) +
  \frac{1}{4} {F}_{mn}(t,t_f+1)+ \frac{1}{4} {F}_{mn}(t+1,t_f+1)
  \label{eq:Fsmear}\\
  &=\frac{B_{mn}}{B_{00}}(1+\bar{c}_{mn}(t,t_f)-\bar{c}_{00}(t,t_f)) 
  \label{eq:Fbar}
\end{align}
The coefficients of contamination terms from the parity partners are
suppressed by factor of $1/2 \sim 1/12$ in $\bar{c}_{mn}$ compared
with $\hat{c}_{mn}$
Hence, once we determine $B_{00}$ from the exponential fitting,
the rest of analysis is just a simple fit of $\bar{F}_{mn}$ to
a constant.
\input{fig_F_Rb}
In Fig.~\ref{fig:F11+rRb}\;\subref{fig:F11}, we show results of
fitting the $\bar{F}_{11}^{B\to D^\ast}$ data to a constant as an
example.
%

%==============================================================
%% fit R
\subsection{Analysis of $R$}
\label{ssec:R}

By construction, \rg{the leading exponential dependence cancels out in 
the double ratio $R$ defined in Eq.~\eqref{eq:R:h_A1}, so the correction
due to the contamination from the excited states is further reduced. 
To further suppress this contamination}, we
take the same linear combination as for $\bar{F}_{mn}$ in
Eq.~\eqref{eq:Fsmear}:
\begin{align}
  \bar R(t,t_f)&\equiv \frac{1}{2}R(t,t_f) + \frac{1}{4}R(t,t_f+1) +
  \frac{1}{4}R(t+1,t_f+1).
  \label{eq:Rbar}
\end{align}
and fit the results to a constant.
In Fig.~\ref{fig:F11+rRb}\;\subref{fig:rRb}, we show results of
fitting \rg{data for $\sqrt{\bar{R}}$, obtained
using the currents improved up to $\mathcal{O}(\lambda^3)$, to a constant.}
\input{fig_cmp_rRb}
In Fig.~\ref{fig:rR}\;\subref{fig:Rsqrt}, we present results for
$\sqrt{R}$ obtained in three different \rg{ways: (i) fit 
$\sqrt{\bar{R}}$ to a constant; (ii) obtain $B_{00}$
by fitting the data at $t_f = 12$ and combine it with results of
$\bar{F}_{mn}$, and (iii)} obtain $B_{00}$ by fitting the data
at $t_f = 13$ and combining it with results of $\bar{F}_{mn}$.
\rg{Results from the first method are given using red crosses, second
  method using blue squares, and the green circle for the third
  method. We find all three are consistent within statistical
  uncertainty.}

\rg{To evaluate the corrections to $\sqrt{R}$ from the improvement
terms in the currents consider the quantity} 
\begin{align}
  [\Delta  R^{1/2} ]_n \equiv \frac{  [R^{1/2}]_n
    -  [R^{1/2}]_{n-1} }{ [R^{1/2}]_{n-1} }.
  \label{eq:DR}
\end{align}
where the subscript $n$ in $[R^{1/2}]_n$ represents results obtained
using the currents improved up to $\mathcal{O}(\lambda^n)$.

In Fig.~\ref{fig:rR}\;\subref{fig:deltaRsqrt}, we present results for
$[\Delta R^{1/2} ]_n$ as a function of $n$ with $n=1,2,3$ \rg{using 
the same convention for the symbols as in
Fig.~\ref{fig:rR}\;\subref{fig:Rsqrt}.}
\rg{Due to a} complicated structure of cancellation in the double ratio,
we do not expect to see a simple scaling behavior in
$\lambda^n$ in $[\Delta R^{1/2} ]_n$.
However, we observe a kind of scaling behavior in $[\Delta B]_n$
defined similarly as in Eq.~\eqref{eq:DR}.
\rg{For a detailed analysis of these behaviors, we refer the reader to 
Ref.~\cite{Park:2017}.}
%

%%%%%%%%%%%%%%%%%%%%%%%%%%%
% Section
%%%%%%%%%%%%%%%%%%%%%%%%%%%
%%%\section{Summary and Plan \label{sec:summary}}

\section*{Acknowledgment}

\begin{acknowledgement}
%
% We thank Jon A.~Bailey for helpful comments and suggestions.
%
\rg{We thank the MILC collaboration for sharing the 2+1+1-flavor HISQ ensembles 
generated by them. }
The research of W.~Lee is supported by the Creative Research
Initiatives Program (No.~2017013332) of the NRF grant funded by the
Korean government (MEST).
J.A.B is supported by the Basic Science Research Program of the
National Research Foundation of Korea (NRF) funded by the Ministry of
Education (No.~2015024974).
W.~Lee would like to acknowledge the support from the KISTI
supercomputing center through the strategic support program for the
supercomputing application research (No.~KSC-2015-G2-002).
The research of T. Bhattacharya, R. Gupta and Y-C. Jang is
supported by the U.S. Department of Energy, Office of Science of High
Energy Physics under contract number~DE-KA-1401020, the LANL LDRD
program and Institutional Computing.
Computations were carried out in part on the DAVID clusters at Seoul
National University.
\end{acknowledgement}
%
%\clearpage
\bibliography{lattice2017}

%%%%%%%%%%%%%%%%%%%%%%%%%%%%%%%%%%%%%%%%%%%%%%%%%%%%%%%%%%%%%%%%%%%%%%%%%%%%%
\end{document}

%% file: macro.tex
\providecommand{\wbar}[1]{\overline#1}

\providecommand{\MeV}{\,\mathrm{MeV}}
\providecommand{\fm}{\,\mathrm{fm}}

\newcommand{\CR}{{\cal R}}
\newcommand{\kcrit}{\kappa_\text{crit}}
\newcommand{\tree}{\text{tree}}
\newcommand{\BtoDst}{\bar B \rightarrow D^\ast \ell \bar \nu}

\newcommand{\epsK}{\varepsilon_{K}}

\newcommand{\rg}[1]{{#1}} % unhighlight Rajan

%% file: fig_kcrit.tex
\begin{figure}
  \sidecaption
  \includegraphics[width=0.55\textwidth]{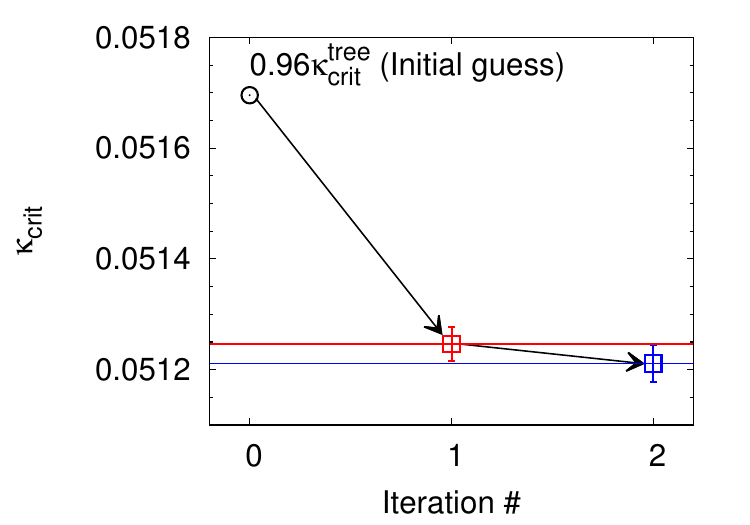}
  \caption{ $\kcrit$ as a function of the iteration number.
    The black circle is the initial guess, the red square is $\kcrit$ after one iteration, and 
    the blue square is the $\kcrit$ after two iterations. }
  \label{fig:kcrit}
\end{figure}

%% file: tab_kc_kb.tex
\begin{table}[h]
%%%  \centering
  \caption{\rg{Results for $\kappa_b$ and $\kappa_c$. The lattice spacing $a$ is set in two ways: using 
    $a_{f_\pi^+} = 0.12520(22)\fm$ from $f_\pi^+$ in~\cite{Bazavov:2014wgs} 
    and $a_{r_1} = 0.1207(11)\fm$ using $r_1$ from~\cite{ Bazavov:2012xda}. The first error is statistical, the
    second is from the experimental error in $M^X$, and the third
    is a systematic error due to the fitting ambiguity.}}
  \label{tab:kc,kb}
  \renewcommand{\arraystretch}{1.2}
  \subtable[charm]{  \resizebox{0.495\textwidth}{!}{
      \begin{tabular}{ l l l }
        \hline\hline
        X & $\kappa_c$ ($a_{f_{\pi^+}}$) & $\kappa_c$ ($a_{r_1}$)
        \\ \hline
        pseudoscalar  & 0.048349(35)(9)(4)& 0.048524(33)(43)(0)
        \\
        vector & 0.048338(62)(11)(0) & 0.048533(59)(48)(1)
        \\
        spin-average & 0.048341(51)(10)(1) & 0.048531(48)(48)(1)
        \\ \hline\hline
      \end{tabular}
    }
    \label{tab:kc}
  }
  \hfill
  \subtable[bottom]{  \resizebox{0.47\textwidth}{!}{
      \begin{tabular}{ l ll }
        \hline\hline
        X  & $\kappa_b$ ($a_{f_{\pi^+}}$)& $\kappa_b$ ($a_{r_1}$)
        \\\hline
        pseudoscalar  & 0.04065(15)(2)(0) & 0.04102(14)(9)(0)
        \\
        vector & 0.04084(18)(2)(1) & 0.04122(18)(10)(1)
        \\
        spin-average  & 0.04079(17)(2)(1) & 0.04117(16)(10)(0)
        \\ \hline\hline
      \end{tabular}
    }
    \label{tab:kb}
  }

\end{table}

%% file: fig_I.tex
%
% Inconsistency parameter
%
\begin{figure*}
  \subfigure[Total view]{\includegraphics[width=0.5\textwidth]
    {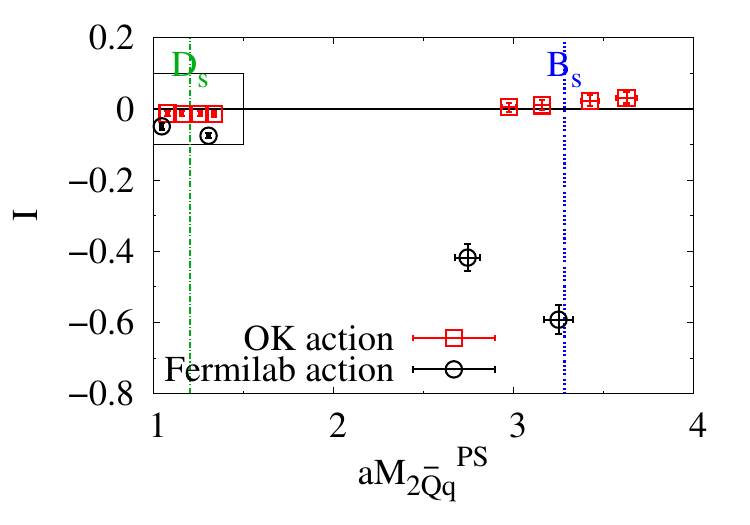}
    \label{fig:I-kcnp}}
  \subfigure[Zoomed-in view]{\includegraphics[width=0.5\textwidth]
    {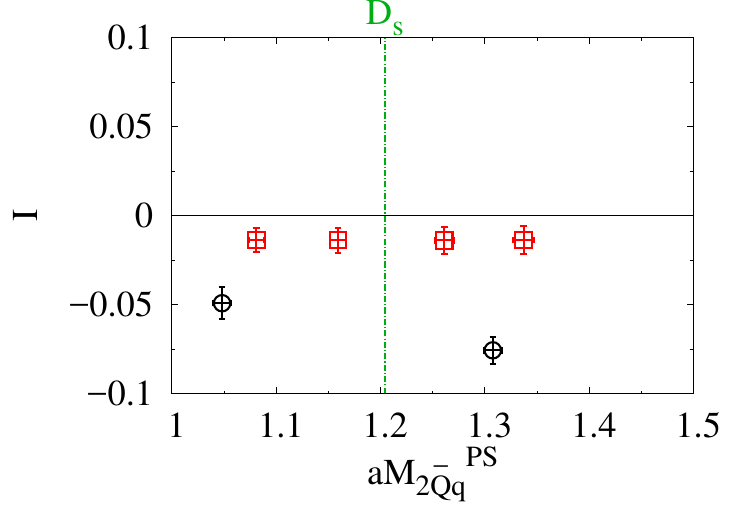}
    \label{fig:I-Ds}}
  \caption{ \subref{fig:I-kcnp} Inconsistency parameter $I$ as a
    function of pseudo-scalar heavy-light meson mass and
    \subref{fig:I-Ds} Zoomed-in view of the box near the $D_s$
    region. Here, we use $\kcrit^\text{NP}$ to measure the 2-point
    meson correlation functions.  The black circles represents results
    obtained using the Fermilab action with the asqtad strange
    quark. For more details, refer to Ref.~\cite{Bailey:2017nzm}. The
    red squares represent results obtained using the OK action with
    the HISQ strange quark.  Vertical dotted lines indicate the
    physical $B_s$ and $D_s$ pseudoscalar mesons.}
  \label{fig:I}
\end{figure*}

%% file: fig_F_Rb.tex
\begin{figure}[htbp]
  \subfigure[$\bar{F}_{11}^{B\to D^\ast}$]{
    \includegraphics[width=0.49\textwidth]{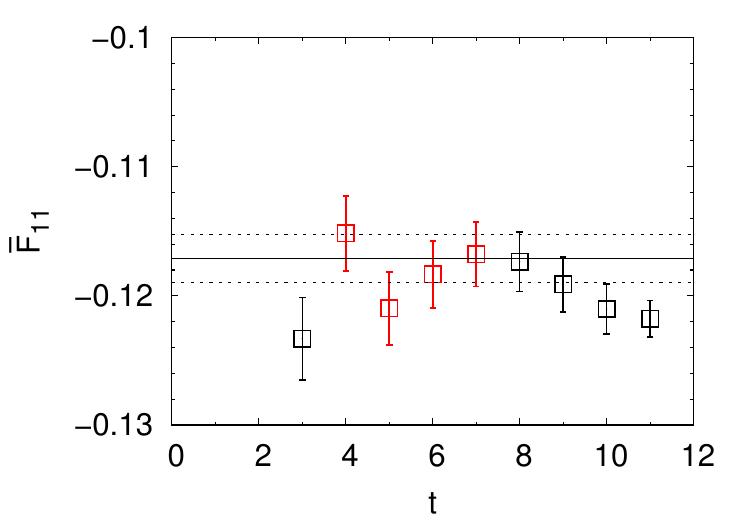}
    \label{fig:F11} }
  \hfill
  \subfigure[$\sqrt{\bar{R}}$]{
    \includegraphics[width=0.49\textwidth]{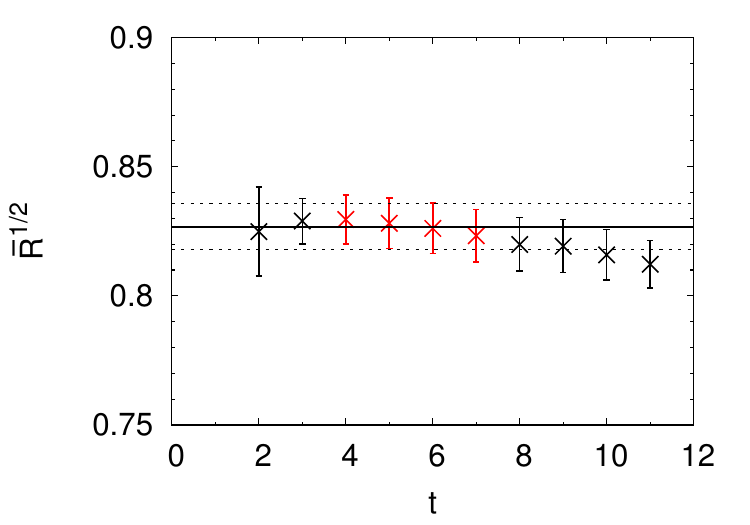}
    \label{fig:rRb} }
  \caption{ \subref{fig:F11} $\bar{F}_{11}^{B\to D^\ast}$ and
    \subref{fig:rRb} $\sqrt{\bar{R}}$ as a function of time.
    Horizontal lines represent results of the constant fit.  The red
    symbols represent those data points used for fitting. We use the
  $\mathcal{O}(\lambda^3)$ improved currents to obtain $\bar{R}$. }
  \label{fig:F11+rRb}
\end{figure}

%% file: fig_cmp_rRb.tex
\begin{figure}
  \subfigure[$\sqrt{R}$]{
    \includegraphics[width=0.49\textwidth]{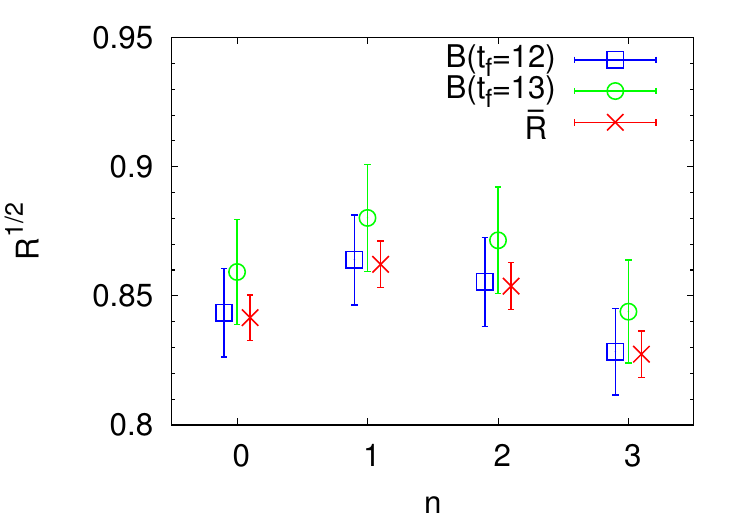}
    \label{fig:Rsqrt}}
  \hfill
  \subfigure[${[}\Delta \sqrt{R}{]}_n$]{
    \includegraphics[width=0.49\textwidth]{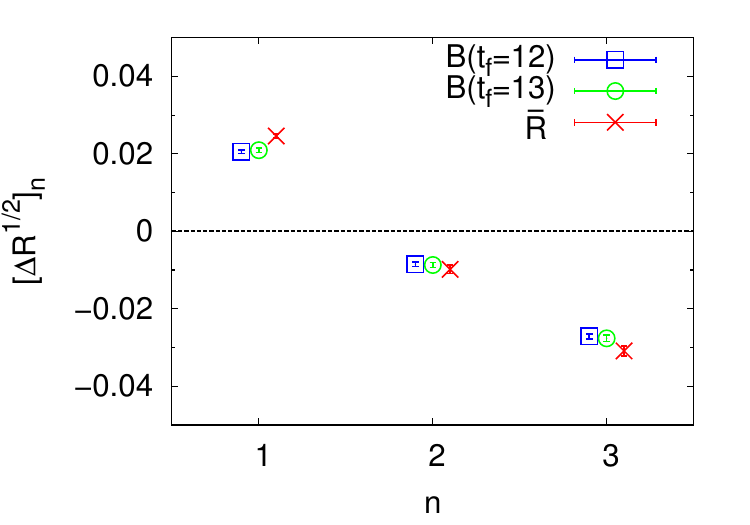}
    \label{fig:deltaRsqrt}}
  \caption{ \subref{fig:Rsqrt} $\sqrt{R}$ as a function of $n$, and
    \subref{fig:deltaRsqrt} $[\Delta \sqrt{R}]_n$ as a function of
    $n$.  Here, $n$ represents the data points obtained using the
    currents improved up to $\mathcal{O}(\lambda^n)$. 
    \rg{Data for $R$ at $t_f=12$ and $t_f=13$ is used to reconstruct
$\bar{R}$.}
}
  \label{fig:rR}
\end{figure}